\documentclass[twocolumn]{aastex63}
\usepackage{amsmath}
\usepackage{amssymb}
\usepackage{cases}
\usepackage{mathrsfs}
\usepackage[flushleft]{threeparttable}
\usepackage{amssymb}
\usepackage{pifont}
\usepackage{amsmath}
\usepackage{epstopdf}
\usepackage{xcolor}
\usepackage[all]{hypcap}
\usepackage{mwe,tikz}
\usepackage{bm}
\usepackage{tabularx}
\usepackage{hyperref}
\usepackage{xspace}
\usepackage{rotating}

\newcommand{\e}{\text{e}}

\newcommand{\p}{\partial}

\newcommand{\fermi}{{\em Fermi}\xspace}

\newcommand{\sw}[1]{\texttt{#1}}

\shorttitle{Magnetar Giant Flare Originated GRB 200415A}
\shortauthors{Vikas Chand et al.}

\begin{document}

\title{Magnetar Giant Flare Originated GRB 200415A: Transient GeV emission, Time-Resolved $\rm E_p~ -~L_{iso}$ Correlation, and Implications}

\email{*: vikasK2@nju.edu.cn, \\ $\dagger$: jjoshi@nju.edu.cn,\\ **: bbzhang@nju.edu.cn,\\ $\ddagger$: manoneeta@iiti.ac.in}

\author[0000-0002-7876-7362]{Vikas Chand$^*$}
\affiliation{School of Astronomy and Space Science, Nanjing
University, Nanjing 210093, China}
\affiliation{Key Laboratory of Modern Astronomy and Astrophysics (Nanjing University), Ministry of Education, China}

\author[0000-0003-3383-1591]{Jagdish C. Joshi$^\dagger$}
\affiliation{School of Astronomy and Space Science, Nanjing
University, Nanjing 210093, China}
\affiliation{Key Laboratory of Modern Astronomy and Astrophysics (Nanjing University), Ministry of Education, China}

\author[0000-0003-4905-7801]{Rahul Gupta}
\affiliation{Aryabhatta Research Institute of Observational Sciences (ARIES), Manora Peak, Nainital-263002, India.}
\affiliation{Department of Physics, Deen Dayal Upadhyaya Gorakhpur University, Gorakhpur 273009, India} 

\author[0000-0003-0691-6688]{Yu-Han Yang}
\affiliation{School of Astronomy and Space Science, Nanjing
University, Nanjing 210093, China}
\affiliation{Key Laboratory of Modern Astronomy and Astrophysics (Nanjing University), Ministry of Education, China}

\author[0000-0001-9868-9042]{Dimple}
\affiliation{Aryabhatta Research Institute of Observational Sciences (ARIES), Manora Peak, Nainital-263002, India.} 
\affiliation{Department of Physics, Deen Dayal Upadhyaya Gorakhpur University, Gorakhpur 273009, India}

\author[0000-0002-4394-4138]{Vidushi Sharma}
\affiliation{Inter University Centre for Astronomy and Astrophysics, Pune, India}

\author[0000-0002-5485-5042]{Jun Yang}
\affiliation{School of Astronomy and Space Science, Nanjing
University, Nanjing 210093, China}
\affiliation{Key Laboratory of Modern Astronomy and Astrophysics (Nanjing University), Ministry of Education, China}

\author[0000-0002-9736-9538]{Manoneeta Chakraborty$^\ddagger$}
\affiliation{DAASE, Indian Institute of Technology Indore, Khandwa Road, Simrol, Indore 453552, India}

\author{Jin-Hang Zou}
\affiliation{College of Physics, Hebei Normal University, Shijiazhuang 050024, China}
\author{Lang Shao}
\affiliation{College of Physics, Hebei Normal University, Shijiazhuang 050024, China}

\author[0000-0002-7555-0790]{Yi-Si Yang}
\affiliation{School of Astronomy and Space Science, Nanjing
University, Nanjing 210093, China}
\affiliation{Key Laboratory of Modern Astronomy and Astrophysics (Nanjing University), Ministry of Education, China}

\author[0000-0003-4111-5958]{Bin-Bin Zhang$^{**}$}
\affiliation{School of Astronomy and Space Science, Nanjing
University, Nanjing 210093, China}
\affiliation{Key Laboratory of Modern Astronomy and Astrophysics (Nanjing University), Ministry of Education, China}
\affiliation{Department of Physics and Astronomy, University of Nevada Las Vegas, NV 89154, USA}

\author{S. B. Pandey}
\affiliation{Aryabhatta Research Institute of Observational Sciences (ARIES), Manora Peak, Nainital-263002, India.}

\author[0000-0002-4371-2417]{Ankush Banerjee}
\noaffiliation{}
\author[0000-0002-4371-2417]{Eman Moneer}
\affiliation{Princess Nourah Bint Abdurhamn University Department of Physics, KSA Riyadh \mbox{84428}~Airport Road}

\begin{abstract}
Giant flares (GFs) are unusual bursts from soft gamma-ray repeaters (SGRs) that release an enormous amount of energy in a fraction of a second. The afterglow emission of these SGR-GFs or GF candidates is a highly beneficial means of discerning their composition, relativistic speed, and emission mechanisms. GRB 200415A is a recent GF candidate observed in a direction coincident with the nearby Sculptor galaxy at 3.5 Mpc. In this work, we searched for transient gamma-ray emission in past observations by \fermi-LAT in the direction of GRB 200415A. These observations confirm that GRB 200415A is observed as a transient GeV source only once. A pure pair-plasma fireball cannot provide the required energy for the interpretation of GeV afterglow emission and a baryonic poor outflow is additionally needed to explain the afterglow emission. A baryonic rich outflow is also viable, as it can explain the variability and observed
quasi-thermal spectrum of the prompt emission if dissipation is happening below the photosphere via
internal shocks. Using the peak energy ($E_p$) of the time-resolved prompt emission spectra and their fluxes ($F_p$), we found correlation between $E_p$ and $F_p$ or isotropic luminosity $L_{\rm iso}$ for GRB 200415A. This supports the intrinsic nature of $E_p$-$E_{\rm iso}$ correlation found in SGRs-GFs, hence favoring a baryonic poor outflow. Our results also indicate a different mechanism at work during the initial spike, and that the evolution of the prompt emission spectral properties in this outflow would be intrinsically due to the injection process.
\end{abstract}


\section{Introduction} 
\label{sec:intro}
SGRs are young, slow-spinning magnetars, exhibiting tens to hundreds of short (duration of ms to s), repetitive bursts in a soft gamma-ray band  \citep{Duncan:1992ApJ392L, Thompson:1995MNRAS}. Their spin periods range from 2 to 12 s and the corresponding spin-down ages vary between $10^3-10^5$ yrs. During their active outburst phases, the magnetars exhibit strong flaring activities spanning a wide range of intensity and durations \footnote{\url{https://staff.fnwi.uva.nl/a.l.watts/magnetar/mb.html}}.
Magnetar flare emission activities are broadly classified into (i) short bursts ($10^{36} - 10^{41}$ $\rm erg ~s^{-1}$) that last for a duration ranging from a few milliseconds to a few seconds, (ii) intermediate bursts ($10^{41} - 10^{43}$ $\rm erg ~s^{-1}$) or (iii) giant flares (GFs) ($10^{44} - 10^{47}$  $\rm erg ~s^{-1}$) lasting for several minutes (see e.g., \citealt{Kaspi:2017ARA}). 
The GFs of SGRs are typically characterized by a short hard initial intensity spike followed by a gradual intensity decay over hundreds of seconds, during which oscillations corresponding to the spin period of the magnetar are observed. The initial intense intensity spike corresponds to a spectrally hard emission with energies greater than 60 keV \citep{Palmer:2005Nature} lasting for typical duration of $\sim 0.1$ s.
The GF events are extremely rare and till date, only three confirmed GFs have been discovered since 1979 from SGRs: SGR 0526--66 \citep{Mazets:1979Nature, Mazets:1982Ap&SS}, SGR 1900+14 \citep{Cline:1998IAUC, Hurley:1999Nature, Kouveliotou:1999ApJ, Mazets:1999AstL}, and SGR 1806--20 \citep{Mereghetti:2005ApJ.624L.105M, Palmer:2005Nature, Hurley:2005Nature, Frederiks:2007AstL}, respectively. There is another possible candidate GF detected SGR 1627--41 \citep{1999ApJ519L151M, Woods:1999ApJ} but not confirmed.
Because of their intense gamma-ray luminosity and spectral characteristics, distant (extra-galactic) SGR-GFs have long been proposed to contribute at least a subset of the observed short Gamma-ray Bursts (sGRBs) \citep{Hurley:2005Nature}. GRB 051103 \citep{Frederiks:2007AstL33, Ofek:2006ApJ.652.507O} and GRB 070201 \citep{Mazets:2008ApJ, Ofek:2008ApJ} in the past have been proposed as candidates of GF-sGRBs. GRB 200415A is the first extragalactic GF candidate observed by \fermi space observatory \citep{Jun_Yang:2020, Zhang:2020arXiv200805097Z}.

The physical mechanism of GFs is still a puzzle despite
many investigations \citep{Thompson:1995MNRAS, Lyutikov:2003MNRAS.346.540L, Parfrey_2013}. The relation between the time-integrated properties of the GFs \citep{Zhang:2020arXiv200805097Z} and in their evolution during a GF (this work) can provide insights into these. Afterglow emissions are another possible electromagnetic counterpart of GFs. Previously, radio afterglow emissions have been detected from two GF sources \citep{Frail:1999Natur, Gaensler:2005Natur, Cameron:2005Natur}. For GRB 200415A, emission in \fermi-Large area telescope is reported \citep{Omodei:2020GCN.27597....1O}. In the \fermi era, GRB 200415A is the first GF candidate, and therefore the observed LAT emission would be the first detection of GeV radiation from magnetars. Prior to this source, GeV radiation is not detected from magnetars, and only upper limits on flux $\sim \rm 10^{-12} - 10^{-11} erg ~cm^{-2}~ s^{-1}$ are known \citep{Li_2017}. Hence, it forms an exquisite opportunity to look into the energetic and origin of this emission also in conjunction with observed prompt emission properties.

\section{Analysis}

\subsection{Prompt Emission}
The isotropic equivalent energy of the GF and peak luminosity, assuming its association with the Sculptor galaxy (NGC 253 at 3.5 Mpc), are estimated as $\rm E_{\gamma,\rm iso}=1.36_{-0.13}^{+0.14}\times 10^{46}$ erg and $\rm L_{\gamma,\rm p,iso}=1.62_{-0.16}^{+0.21}\times 10^{48}$ $\rm erg\ s^{-1}$, respectively \citep{Jun_Yang:2020}. Therefore, the radiation luminosity for the time-integrated duration of 0.2 s is $\sim 7 \times 10^{46}$ $\rm erg ~s^{-1}$. 
In the time-dependent spectral analysis by \cite {Jun_Yang:2020}, for interval -5 -- 120 ms, blackbody (BB), multicolor-blackbody (mBB) or a quasi-thermal spectrum is a preferred fit. We further resolved their second time-bin where the best fit was mBB based on signal to noise ratio of $\sim20$. The spectral parameters of the modeled powerlaw with an exponential cutoff (CPL) are presented in Table \ref{tab:spec}.

\subsection{The Fermi/LAT Observation of GRB 200415A}
\label{sec:obser}
We retrieved the photon event data files and the spacecraft history files (pointing and livetime history) from the LAT data server\footnote{\url{https://fermi.gsfc.nasa.gov/cgi-bin/ssc/LAT/LATDataQuery.cgi}}
at the updated localisation of the LAT observations J2000 RA, Dec = 11.07, -25.02 degrees \citep{Omodei:2020GCN.27597....1O}. 
The data are obtained in a spatial radius of $40^\circ$ and energy range of 100 MeV to 300 GeV. We selected a $12^\circ$ region of interest centred at the burst location and also constrained the zenith angle to $100^\circ$ to avoid contamination from the Earth's limb. We selected the cleaner "\sw{Pass 8 source}" class (\sw{evclass} = 128 and \sw{evtype} = 3) with response function \sw{P8R3\_SOURCE\_V2}. The corresponding model for extragalactic diffused gamma-ray background \sw{iso\_P8R3\_SOURCE\_V2.txt}
is used and for the galactic contribution, diffused gamma-ray emission is estimated using the official galactic interstellar emission model \sw{gll\_iem\_07.fits}.

The analyses are carried out with the following objectives (1) Ascertain the detection of a transient emission, for this purpose we choose the data in a relatively smalled interval of 10,000 s before the trigger-time, and from the trigger-time to 1000 s. (2) Study the nature of the transient emission, the data are analysed between 5 days prior to the trigger-time ($\rm T_0$) to 5 days after. (3) After this analysis, we aim to confirm if there is any other transient emission before this emission in the past observations of \fermi-LAT. A computationally less expensive approch is followed. The data is first analysed in an interval of a bin size of six months, which is then used to extrapolate to set a reference limit on flux, and the analysis is finally carried out in an interval of bin size of 2 days for 11.5 years of data.

\begin{figure}
\centering
\includegraphics[scale=0.6]{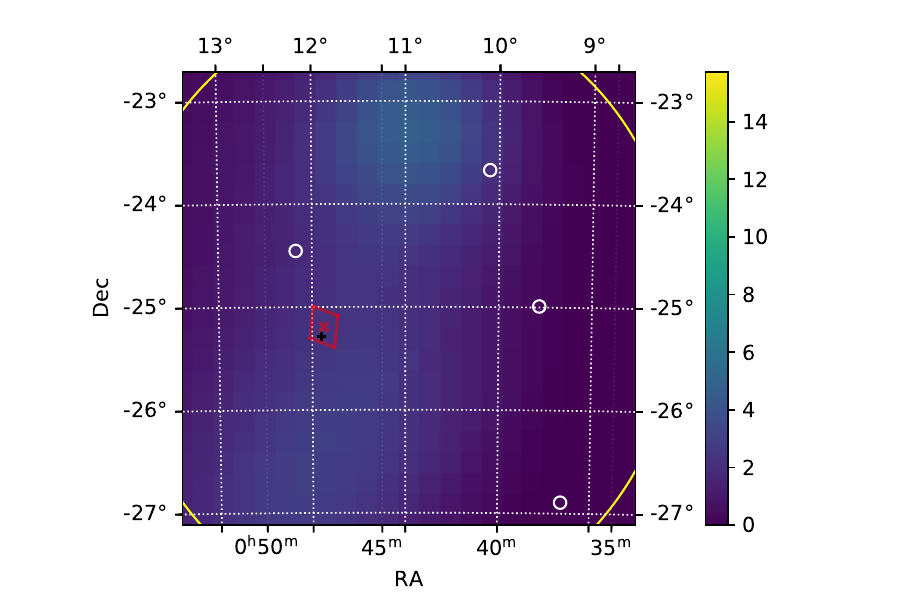}
\includegraphics[scale=0.6]{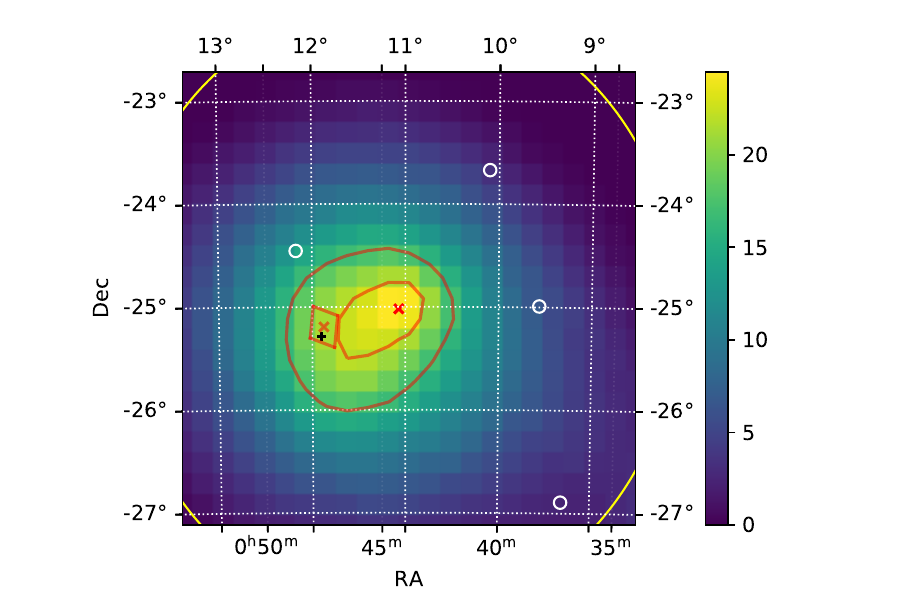}
\includegraphics[scale=0.6]{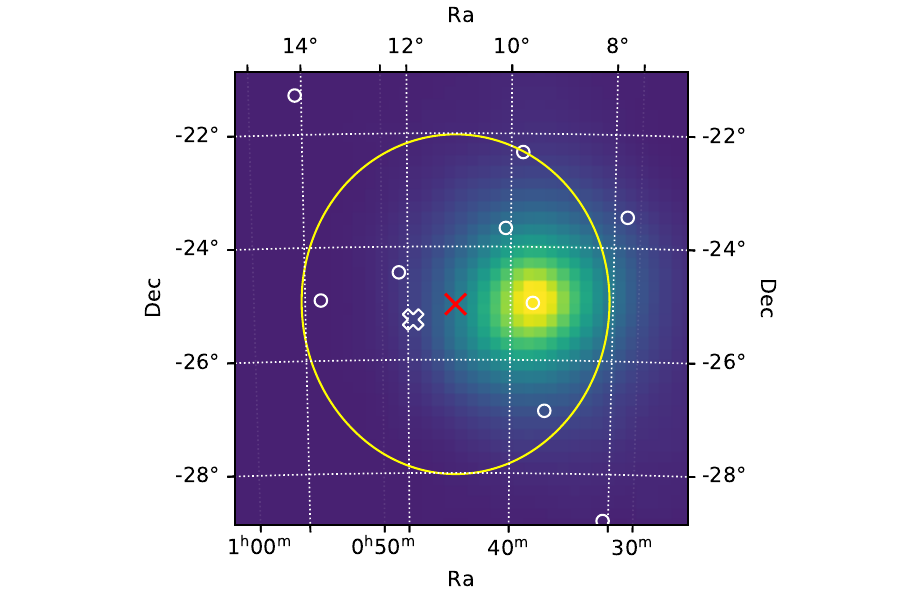}
\caption{The test statistic map (TS) in a region $16^{\circ} \times16^{\circ}$ region at $0.2^{\circ}$ resolution (zoomed for better visualization). The color-bars on each image represent the TS values.
The IPN triangulation localisation-region of the GRB is bounded by red lines centred at the cross. The 
GRB is associated to the Sculptor galaxy (NGC 253) marked in a plus sign. 
\emph{Top}: The TS map from -10000 to 0 s. There is no significant emission within the IPN triangulation region during this interval.
\emph{Middle}: TS 
map from 0 - 1000 s. A maximum TS value of 26 is obtained at the GRB location signaling a $\sim5~\sigma$ detection. The 68\% and 90\% confidence levels are plotted in red contours. 
Other known sources from the \fermi-LAT fourth source catalog are marked by open circles. A bigger circle of $3^\circ$ radius centred on the updated LAT localisation is shown for a reference.
\emph{Bottom}: The location of the many transient detections. The signal with significance of $\approx$ 15 is shown.} 
\label{fig:LAT_detection}
\end{figure}

We performed unbinned likelihood analysis and have shown the residual test-statistic map in Figure \ref{fig:LAT_detection} without contribution from this source. In a short duration of 500 s and 1000 s, the analyses show that the GRB is detected with TS value of 27 and 26, respectively. We plotted the test statistic map (TS) in a region $16^{\circ} \times16^{\circ}$ region at $0.2^{\circ}$ resolution (zoomed for better visualization). The color-bars on each image represent the TS values.
The 
GRB is associated with the Sculptor galaxy (NGC 253). 
A maximum TS value of 26 is obtained at the GRB location from 0 - 1000 s duration signaling a $\sim5~\sigma$ detection. Confidence levels (68\% and 90\%) are plotted.  
Other known sources from the \fermi-LAT fourth source catalog are marked by open circles.

\begin{figure*}
\centering
\includegraphics[scale=0.3]{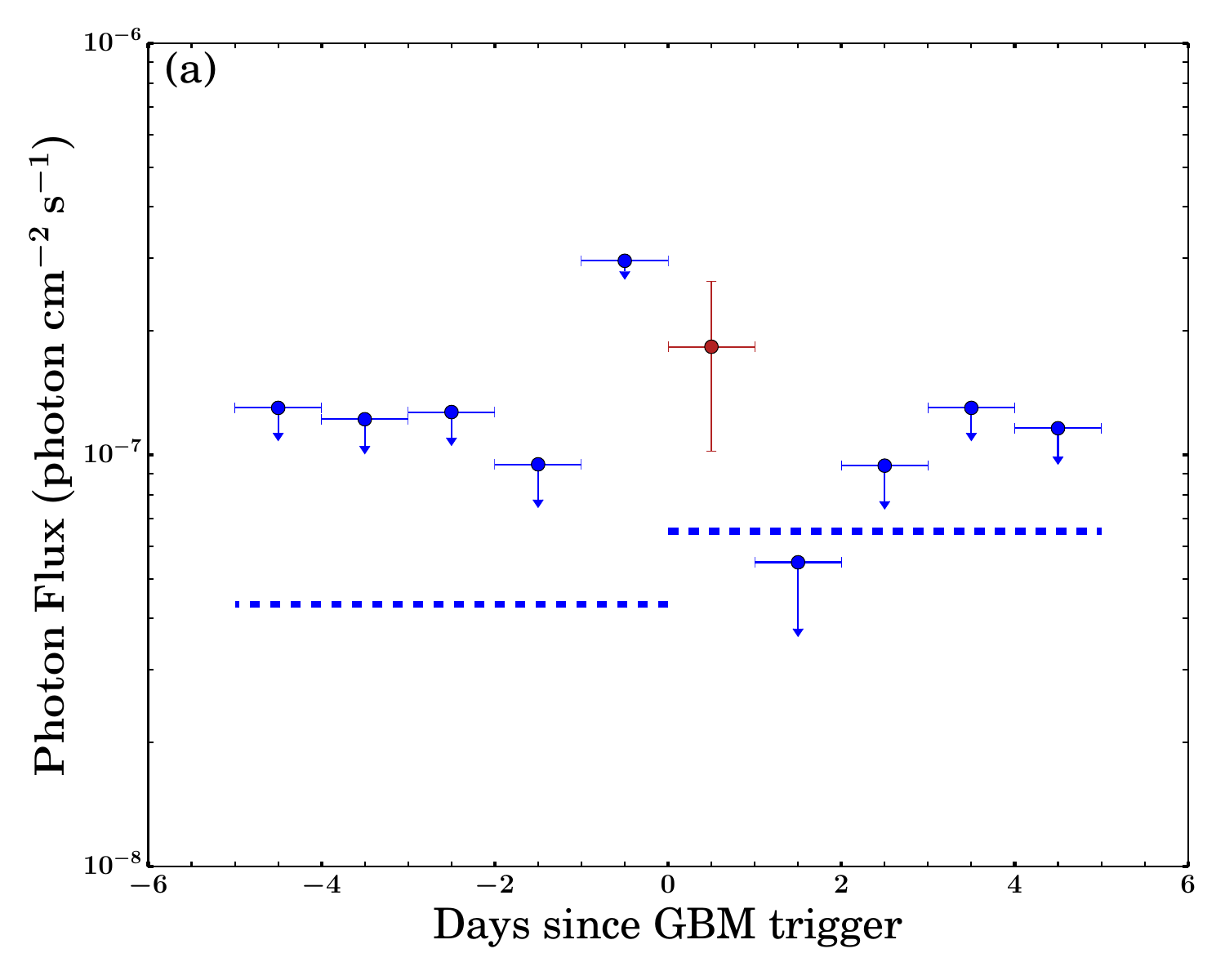}
\includegraphics[scale=0.3]{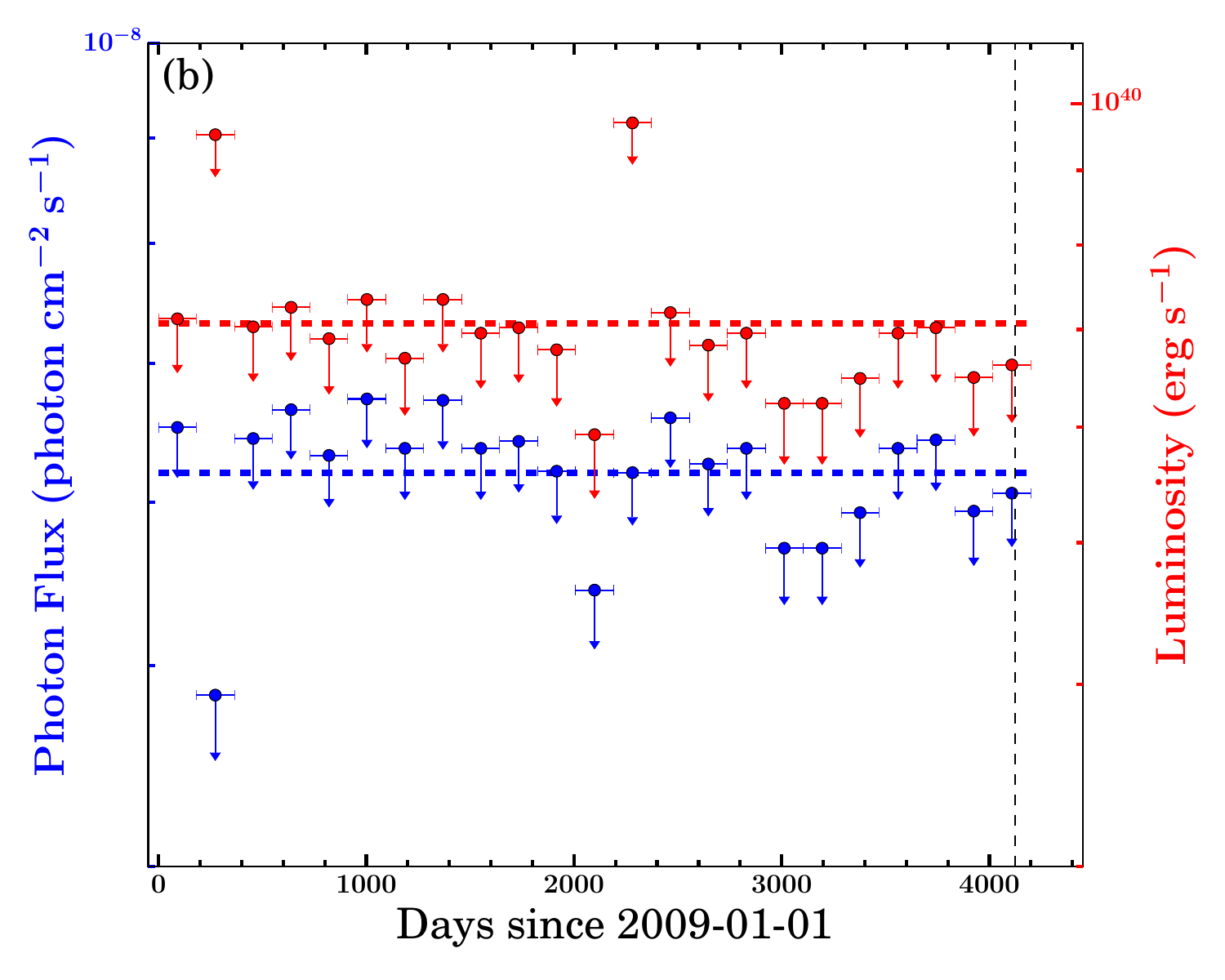}
\includegraphics[scale=0.3]{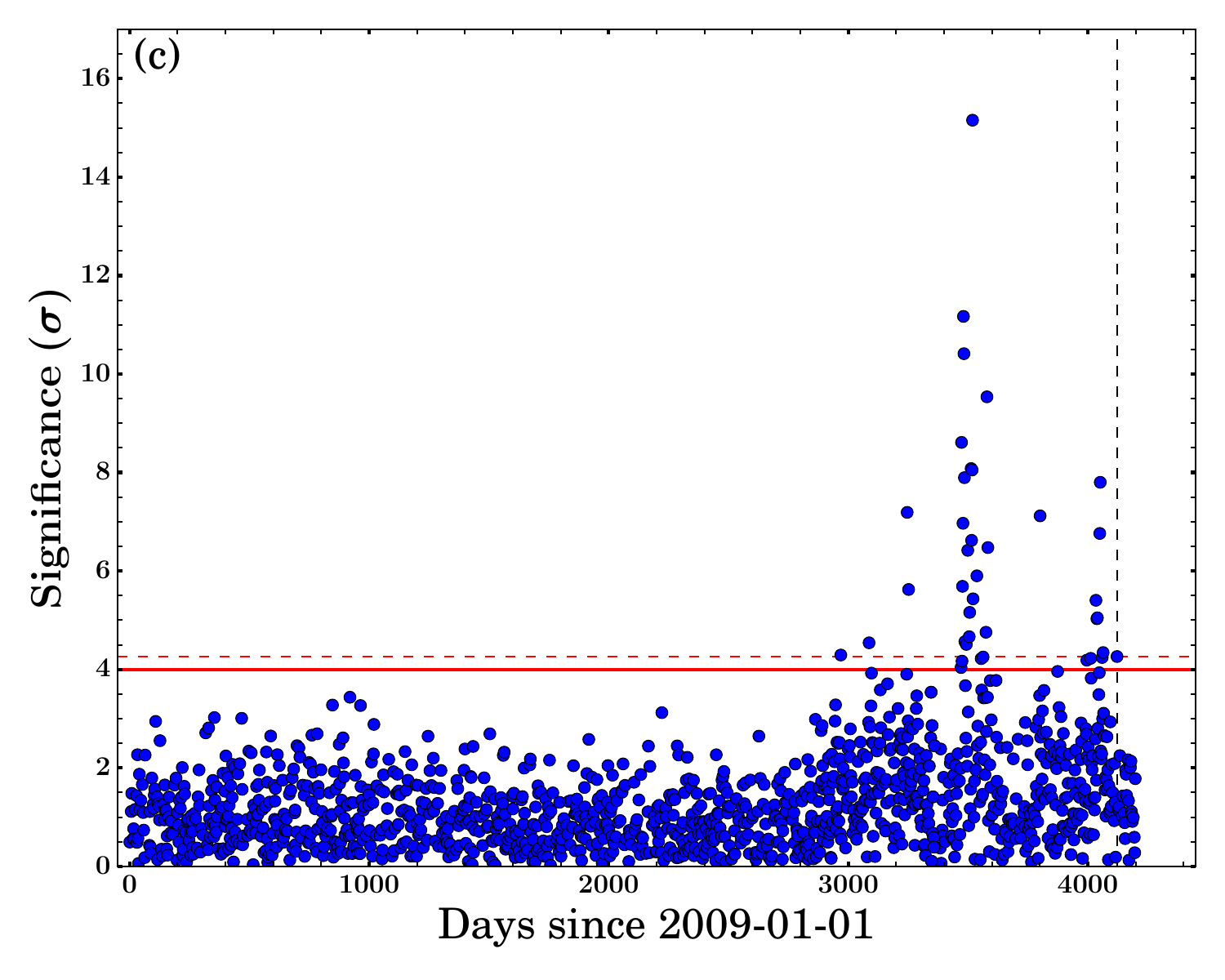}
\includegraphics[scale=0.3]{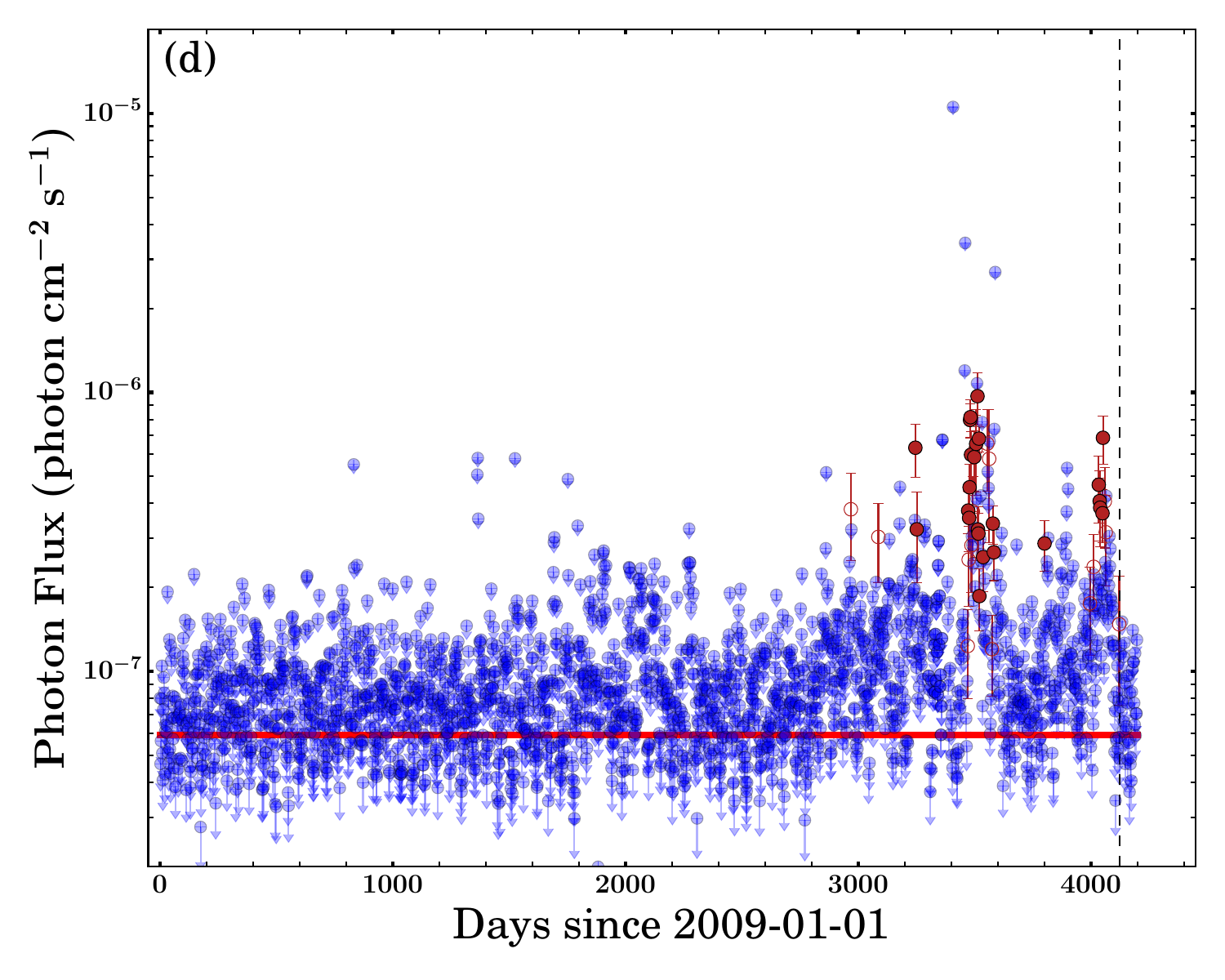}
\caption{(a) Photon fluxes for GRB 200415A from T0-5 days to T+5 days with 1-day bin size. The horizontal blue dashed lines correspond to the total fluxes limit for combined bins from 1 to 5 days before and after the trigger-time. (b) The photon fluxes (blue) and corresponding luminosities (red) upper limits from LAT observations of GRB 200415A location with a half-year bin. The blue and red horizontal dashed lines show the averaged values of the 11.5 years upper limits of photon fluxes and luminosities, respectively. (c) Distribution of significance with 2 days temporal binning. The solid red line shows the TS equal to 16 and the dashed line indicates to TS of the source during bin after trigger-time. (d) \fermi LAT flux light curve for 11.5 years of observations in the direction of GRB 200415A. Blue circles show the upper limits (TS $<$ 16) and firebrick circles show the detection of high energy photons (TS $>$ 16). Open and filled firebrick circles show the TS values 16 $<$ TS $<$ 25 and TS $>$ 25, respectively. The red solid line shows the 2 days photon flux upper limit extrapolated from the average value of the 11.5 years upper limits \citep{2019ApJ...875L..19Y}. The vertical black dashed lines in (b), (c), and (d) denote \fermi GBM trigger-time.}
\label{fig:LAT_1Day}
\end{figure*}

On a scale of 1 day, the photons are still sparsely distributed and thus unbinned likelihood analysis is 
performed using \sw{gtlike}. The GRB contribution to the observed statistics is evaluated using a \sw{powerlaw2}\footnote{\url{https://fermi.gsfc.nasa.gov/ssc/data/analysis/scitools/source\_models.html\#PowerLaw2}} model, and flux (photon and energy flux) is calculated in 0.1 - 10 GeV energy range. 
The upper limits are obtained by assuming a spectral shape with an index equal to -2. LAT detected high energy photons in the first temporal bin (firebrick colored data-point in Figure \ref{fig:LAT_1Day}(a)) after the trigger-time.
To set a reference limit on flux from the direction of the observed LAT emission, we also analysed the 11.5 years of \fermi-data using binned likelihood analysis in bin-sizes of six months. The spectral parameters of the models for the sources within $3^\circ$ centred on the GRB position (as marked by the yellow circle in Figure \ref{fig:LAT_detection}) are kept free for the likelihood analysis. The flux upper limits and corresponding luminosity are plotted in Figure \ref{fig:LAT_1Day} (b).

We binned the data since 2009-01-01 UTC in a bin size of 2 days. The known point sources in ROI are  also included\footnote{user contributed software \url{https://fermi.gsfc.nasa.gov/ssc/data/analysis/user/python3/make4FGLxml.py}} and spectral parameters are fixed to the values reported in the fourth \fermi source catalog \citep{Abdollahi:2020ApJS}. The significance of the emission $\sim \sqrt{TS}$ is plotted in Figure \ref{fig:LAT_1Day} (c). The flux averaged over the six months upper limits (blue dashed line in Figure \ref{fig:LAT_1Day} (b)) is extrapolated (e.g. as in \cite{Yang_2019}) to a bin size of 2 days and shown in red line in Figure \ref{fig:LAT_1Day} (d)).
We also show the TS map at the location of the many detections in the bottom panel of Figure \ref{fig:LAT_detection}. 
Other detections with TS $>$ 16 represented in Figure \ref{fig:LAT_1Day} 
in firebrick color are also at this location and therefore only one with the largest significance (significance $\approx$ 15) is shown for a reference. Here, the IPN localisation area is not encircled by the confidence contours as in the case of the emission observed $\sim19$ s after the Fermi trigger.

\section{Results}
An injected GF gamma ray luminosity of $\rm L_{\gamma} \sim 7 \times 10^{46}$ $\rm erg ~ s^{-1}$ near the neutron star (radius $\rm R_0 \sim 10^6$ cm), produces a fireball, which expands due to its own radiation pressure. The optical depth for $\gamma-\gamma$ interaction is $\rm \tau_{\gamma\gamma} \gtrsim E\sigma_T/(4\pi e_\gamma R_0 ct) \approx 1.36\times 10^{11} E_{46.13} R^{-1}_{0,6} t^{-1}_{-0.7}$. Here, we used notation $\rm X = 10^n X_n$, and $\rm \sigma_T$ is the Thompson cross-section, $\rm t \sim 0.2 t_{-0.7}$ s is the duration of the prompt emission, and $e_\gamma \sim 900$ keV is the mean energy obtained from time-integrated spectral fit. This huge optical depth creates a radiation and $\rm e^{\pm}$ pair dominated plasma, with initial temperature $\rm T_0 \approx (E/4 \pi R^2 \sigma t)^{1/4} \approx 270{\rm E^{1/4}_{46.13} R_{0,6}^{-1/2} t^{-1/4}_{-0.7}~ keV}$ \citep{Paczynski:1986ApJ,Nakar:2005ApJ.635.516N}. 
The $Lorentz$ factor of this plasma increases with radius $\Gamma \propto \rm R$ and comoving temperature decreases as inversely proportional to radius $\rm T \propto R^{-1}$ \citep{Goodman:1986ApJ, Paczynski:1986ApJ, Shemi:1990ApJ,  Duncan:1992ApJ, Piran:1993MNRAS, Meszaros:1993ApJ181M, Katz:1996ApJ305K}. During this evolution, at some stage, pair production is negligible and we can estimate number of pairs $\rm N_{\pm}  \approx  4\times 10^{44} E_{46.13}^{3/4} R_{0,6}^{-1/2} t_{-0.7}^{1/4}$  \citep{Nakar:2005ApJ.635.516N}. Even after this stage, photons are coupled to the 
pairs through scattering, which accelerate the pair plasma to a bulk $Lorentz$ 
factor $\Gamma_{\pm}$. The bulk $Lorentz$ 
factor and kinetic energy of the pair plasma is given by $\rm \Gamma_{\pm} = (E \sigma_T / 4 \pi c^3 t m_e R_0)^{1/4} \approx  \rm 620 E_{46.13}^{1/4} \rm R_{0,6}^{-1/4} \rm t_{-0.7}^{-1/4}$ and $\rm E_{\pm} = N_{\pm} \Gamma_{\pm} m_e c^2 \approx 1.9 \times 10^{41} \rm{erg}$, 
respectively \citep{Nakar:2005ApJ.635.516N}.

For the afterglow observed by \fermi-LAT in the energy range 0.1-10 GeV, during 0-1000 s, we calculate energy flux, which is $\sim (3.78 \pm 2.24) \times 10^{-09}$  $\rm{erg~cm^{-2}~s^{-1}}$. This component has the total energy $\rm E_{LAT} \approx 5.53 \times 10^{45}$ erg, 
however, the kinetic energy left in pair plasma is $\sim 10^{-5}$ of the energy of GeV emission. This also infers that the observed LAT afterglow can not be powered by ejecta with pure pair fireball composition.

In addition to radiation and $\rm e^{\pm}$ pairs, if the fireball is also loaded with baryons, having baryon injection rate $\rm \dot M$, and total burst luminosity  $\rm L_0 = L_{\gamma} /\xi_{\gamma}$, where $\xi_{\gamma}$ is the conversion efficiency of total burst luminosity into gamma rays. The outflow evolution is parameterized in terms of the dimensionless entropy $\eta = \rm L_0/ \dot{M} c^2$ \citep{Shemi:1990ApJ, 2005ApJ6331013I}. We will consider two possible loading cases, (i)  baryonic poor (BP) outflow, when $\eta > \eta_{*}$, where, $\rm \eta_{*}= (L_0 \sigma_T /4\pi m_p c^3 R_0)^{1/4}=91 L_{0, 46.8}^{1/4} R_{0,6}^{-1/4}$ is the critical entropy \citep{2000ApJ530292M}. In the BP case, the photosphere is in an accelerating phase and photospheric radius is below the saturation radius \citep{2000ApJ530292M}. 
The observed temperature is $\rm T_{ph} = T_0 \sim \rm 270 ~keV$ and quasi-thermal 
 emission has peak at $\sim$900 keV. 
 The emission is radiated away from the photospheric radius and the 
 final $Lorentz$ factor would be $\Gamma_f = \eta_* \approx 90$. From the 
 observed afterglow, and the energy remaining in the baryons, we have $\rm \eta = \xi_L (E/E_{LAT})\times \eta_*$, where $\xi_L$ is the 
 efficiency of conversion of the kinetic energy ($\rm E_K$) into the GeV afterglow. 
 
Since, $\xi_L < 1$, we find $\eta < 227$.
Using this $\eta$ we can constrain the baryonic load, and obtain a value of $\rm M > E/\eta c^2 \approx 6.65 \times 10^{22}$ gm. 
Using $\eta > \eta_*$, we also have an upper limit on baryonic load $\rm M < 1.66 \times 10^{23}$ gm. In the BP case, therefore, using the observed afterglow we constrain the baryonic load to be $\rm 6.65 \times 10^{22} ~ gm < M < 1.66 \times 10^{23}$ gm.

The afterglow discussed here is produced in a forward shock when the ejecta inevitably penetrates through the circumburst medium. At the deceleration time of the external forward shock, the $Lorentz$
factor is one half of the initial $Lorentz$ factor. Assuming peak time to be the observed time when the first photon (probability $\geq$ 0.9) from the source is received. Using the initial $Lorentz$ factor we can constrain the density of the ambient medium \citep{Sari:1999ApJ}.
For $\Gamma_0 \sim \rm 45 E_{K, 47}^{1/8} n_0^{-1/8} t_{\gamma, 2}^{-3/8}$, and $\eta_* = 91$, $t_{\gamma} \sim t_{start}$ = $\rm t_{i,obs} \sim 19$ s (time when first photon with probability $\geq$ 0.9 is received) and using $\rm E_K = (\eta_*/\eta)E_{\gamma, iso}$, $\Gamma_0 = \eta_*$, we found $n < 4 \times 10^{-4} {\rm cm^{-3}}$. Here we have assumed, peak time to be the start of the LAT emission. The density would be lower, if peak occurs later than this. 

The other case is (ii) baryonic rich (BR) ejecta, here the photosphere is in a coasting phase \citep{Nakar:2005ApJ.635.516N}.
The emission observed in \fermi-GBM has a minimum variability timescale ($\rm \sim 2 ~ms$ \citealt{Jun_Yang:2020}), which is one of the extreme values when compared to a sample of GRBs \citep{Jun_Yang:2020}.
Thermal emission from internal shocks can arise if it occurs below the photosphere \citep{Rees_2005}. In such a case for the kinetic energy left in the ejecta can be $\sim 10$ times the radiation energy and a jet configuration or atypical parameters are required to explain the afterglows (e.g., as in the case of  SGR 1806-20 \cite{Ioka:2005ApJ}). The temperature/peak energy tracks the photons flux \citep{Jun_Yang:2020}. Such an evolution can be a resultant of many multiple superimposed pulse evaluations \citep[e.g. as discussed in reference to GRBs by][] {Preece:2016ApJ}. 

\begin{table}
\scriptsize
\label{tab:spec}
\centering
\caption{Time-dependent spectral analysis results based on GBM observations, using CPL model.}
\begin{tabular}{|c|c|c|c|c|}
\hline
Sr. no. &Intervals &  $\Gamma_{\rm ph}$  & $\rm E_{\rm p}$,	& Flux 	\\	
& (s) 	&  &  (keV) & ($10^{-4}$ erg cm$^{-2}$ s$^{-1}$)  \\ 
\hline
1 & (-0.005, -0.003) & 0.36$_{-0.31}^{+0.29}$ & 393.53$_{-38.06}^{+72.77}$   & 1.99$_{-0.39}^{+0.50}$ \\

$2^{a}$ & (-0.003, -0.0024) & 1.54$_{-0.52}^{+1.02}$  &  271.68$_{-27.83}^{+44.14}$   &  2.44$_{-0.58}^{+0.79}$  \\

$3^{a}$ & (-0.0024, -0.001) & 0.16$_{-0.23}^{+0.32}$  & 726.84$_{-118.88}^{+146.29}$  & 3.93$_{-1.06}^{+1.26}$ \\

4 &(-0.001,  0.001) & -0.00$_{-0.16}^{+0.26}$ & 1688.27$_{-224.37}^{+304.76}$  & 6.61$_{-1.66}^{+2.19}$  \\

5 &(0.001, 0.005) & 0.60$_{-0.26}^{+0.46}$ & 857.35$_{-132.59}^{+134.64}$  & 2.06$_{-0.56}^{+0.65}$  \\

6 &(0.005, 0.010) & 1.67$_{-0.68}^{+0.88}$  & 847.03$_{-121.32}^{+198.13}$  & 1.12$_{-0.36}^{+0.48}$  \\

7 &(0.010, 0.020) & 0.53$_{-0.20}^{+0.30}$ &  907.19$_{-89.61}^{+99.54}$ & 1.16$_{-0.20}^{+0.26}$  \\

8 &(0.020, 0.040) & 0.63$_{-0.20}^{+0.38}$ & 743.53$_{-75.94}^{+74.67}$  & 0.55$_{-0.10}^{+0.11}$  \\

9 &(0.040,  0.080) & 0.31$_{-0.16}^{+0.23}$ & 676.90$_{-58.29}^{+66.68}$  & 0.38$_{-0.05}^{+0.07}$  \\

10 & ( 0.080, 0.120) & 0.65$_{-0.35}^{+0.52}$ & 374.23$_{-46.83}^{+63.24}$   & 0.09$_{-0.02}^{+0.03}$  \\


\hline
\end{tabular}
\footnotesize{a: newer bins}
\end{table}

We note that very recently \citep{Zhang:2020arXiv200805097Z} have discussed the magnetar GF origin of the emission in GRB 200415A. 
For the magnetar GF candidates, they have found a correlation similar to the Amati correlation in GRBs. Further, they discussed the outflow composition and energetics using the standard analysis (e.g., in reference to GFs \citealt{Nakar:2005ApJ.635.516N, Ioka:2005ApJ, Dai_2005, Wang_2005}).
They ruled out BR outflow based on the relation between the observed temperature and isotropic energy.

Such a correlation, if intrinsic, should also be present between the considered observables in the time-dependent data (in case of GRBs, e.g., \citealt{Fronter2012ApJ754}). We find that hints of such a correlation
between the peak energy $\rm E_p$ and isotropic luminosity $\rm L_{iso}$ is present with Pearson linear correlation coefficient r = 0.67, and p-value = 0.03;

\begin{equation} 
\label{eq:Ep_Liso}
\rm \left(\frac{E_{p}}{1 keV}\right) \approx 619_{-87}^{+93} \left(\frac{L_{iso}}{10^{47} erg~s^{-1}}\right)^{0.23 \pm 0.10}.
\end{equation}

In another form, we have $\rm log(E_{p,0}) \sim 2.8 + (0.23 \pm 0.10) log(\rm L_{iso, 47})$ which is consistent with their results ($\rm E_p \sim E_{iso}^{1/4}$). The scaling relation between $E_p$ and peak flux $F_p$ is

\begin{equation}\label{eq:Ep_flux}
\rm \left(\frac{E_{p}}{1 keV}\right) 
\approx 676_{-92}^{+98} \left(\frac{F_p}{10^{-4} erg~ cm^{-2}~ s^{-1}}\right)^{0.23 \pm 0.10}
\end{equation}. 

For a BP outflow, the temperature observed is the photospheric temperature ($\rm T \sim T_0$). This implies the observed evolution in the case of this source is intrinsic to the injection process rather than the dynamics of the outflow. Alternatively, multiple thermal shells with varying temperatures might be injected from the central source. This is also evident from the observed BB spectrum in bins of size $\sim 4$ ms or quasi-thermal spectrum in larger bins. The dispersion in the correlation can arise in such a case from other parameters, which may vary within injections, such as radius $\rm R_0$ at which injection occurs. The first three points in the time-dependent spectra during -0.005 to -0.001 s are farther from the fitted correlation (assuming an uptrend is more evident). If we exclude these three data-points, both the correlation and significance are strengthened. The Pearson linear correlation coefficient is r = 0.96 (p = 0.0008), and linear correlation between the logarithmic values of the $\rm E_p ~ \& ~ L_{iso}$ is 0.97 (0.0003). The new relation is ($\rm E_{p, 0} \approx 752_{-35}^{+35} L_{iso, 47}^{0.31 \pm 0.04}$).

\begin{figure}
\centering
\includegraphics[scale=0.37]{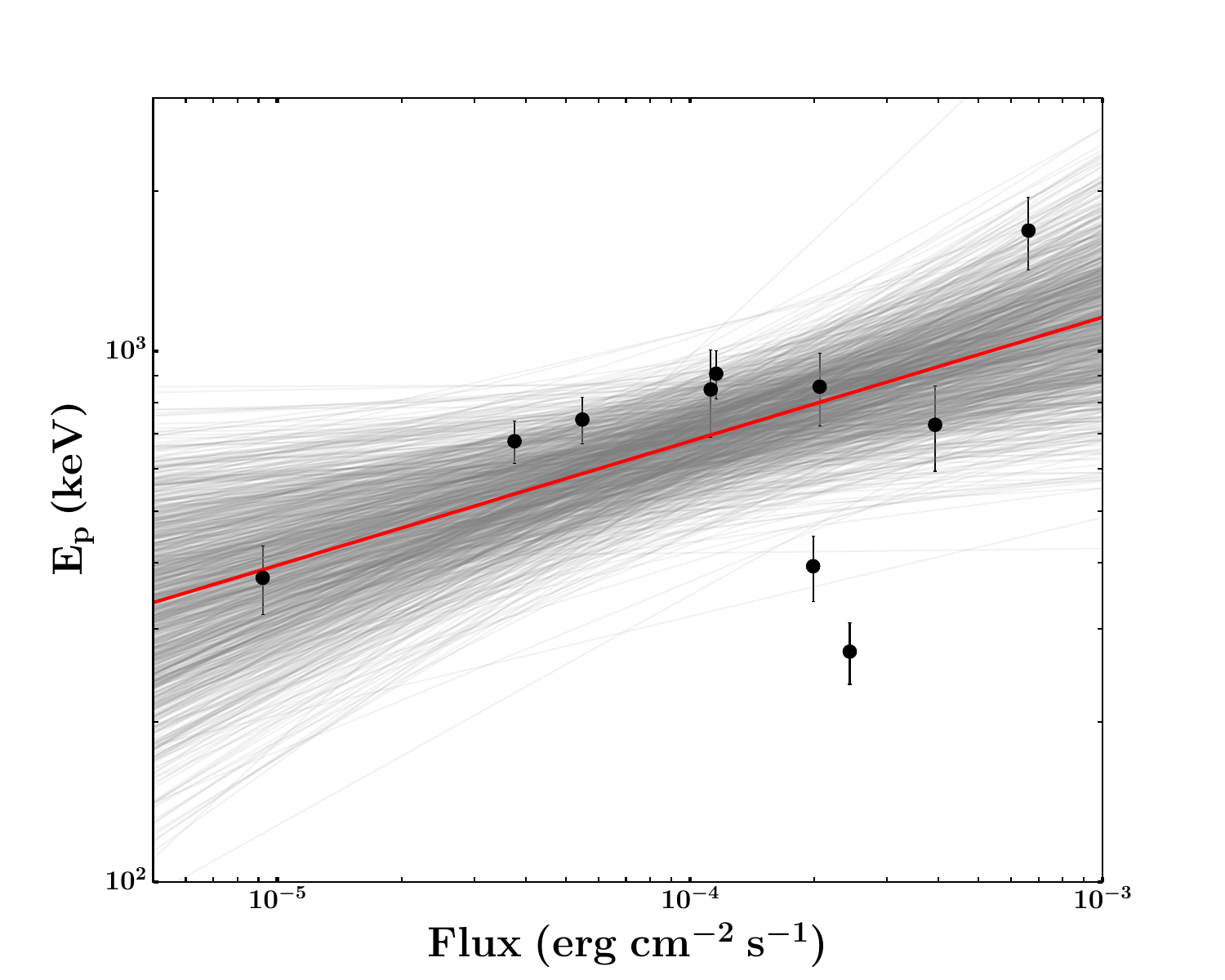}
\includegraphics[scale=0.65]{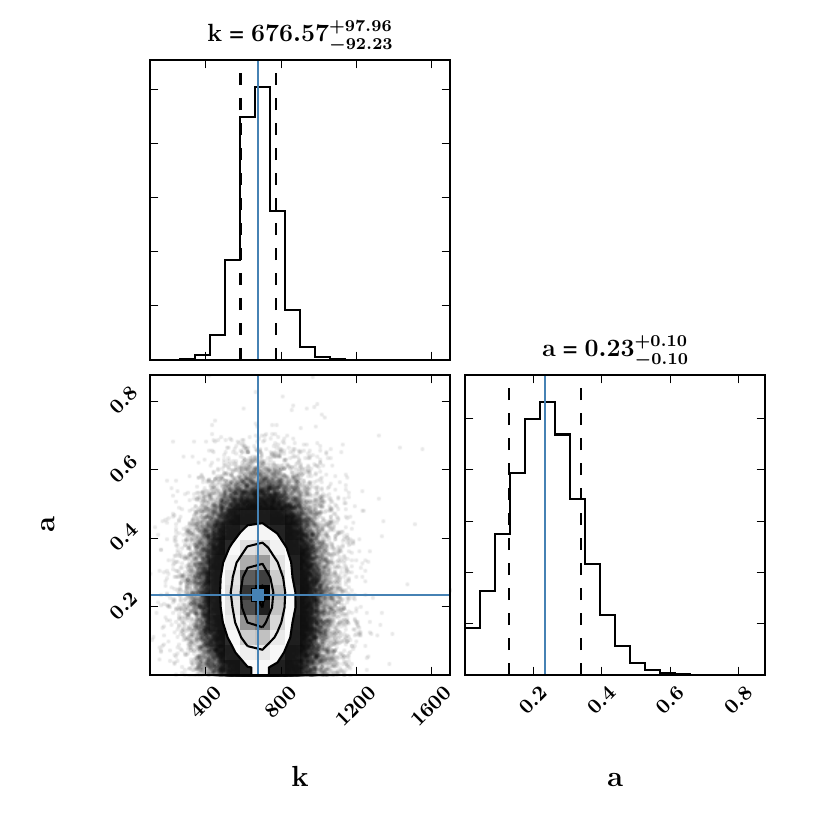}
\caption{{\em Top panel:} Time-dependent $\rm E_p - F_p$ correlation for GRB 200415A. The red line shows the best fit. {\em Bottom panel:} Corner plot shows the results obtained from MCMC simulation for a simple power-law model. a and k represent the index and norm of the powerlaw relation, respectively.}
\label{fig:Ep_Liso_correlation}
\end{figure}

\section{Discussion}

We have systematically studied the afterglow emission in the case of the GF candidate GRB 200415A. No other transient emission in the direction of the GRB was detected in the past 11.5 years of \fermi observations than the one detected $\sim19$ s after the GBM trigger.
We also systematically arrive at the conclusions that the high energy emission as observed in \fermi-LAT is produced in a BP fireball when the ejecta runs into the ambient medium. In this scenario we have constrained the baryonic load in the ejecta. We assumed that the ejecta expands into a constant density medium and using the initial Lorentz factor of the ejecta, onset time of the GeV afterglow, and energy of the ejecta, we have constrained the density of the ambient medium. 

The time-averaged peak energy ($E_p$) and isotropic equivalent energy $E_{\rm iso}$, in a sample of SGR-GF candidates, are correlated \citep{Zhang:2020arXiv200805097Z}. We found a time-dependent correlation between $E_p$ and isotropic luminosity ($L_{\rm iso}$) for GRB 200415A (Eq. \ref{eq:Ep_Liso}). The existence of the time-dependent relation between the $E_p$ and $L_{\rm iso}$ emitted by the source similar to the time-averaged correlation found for the GF candidates favours a BP outflow over a BR outflow \citep{Zhang:2020arXiv200805097Z}. Moreover, this also implies that the observed evolution in the peak energy is intrinsic to the injection process. Finally, our results show that the correlation is tighter if we exclude the initial 4 ms (first 3 time intervals in Table \ref{tab:spec}) implying a different emission mechanism at work during this period.

The physical mechanisms responsible for the GFs have eluded a complete understanding despite many investigations \citep{Thompson:1995MNRAS, Thompson_2001, Lyutikov:2003MNRAS.346.540L, Parfrey_2013, Takamoto:2014ApJ}. The spectral evolution and the variability during a GF indicate the complex nature of the central source. 
This can be probed by a search for other observed features e.g. QPOs that originate in a magnetic reconnection scenario and seismic modes in the magnetar crust developed during the GF and aided by the intense magnetic field \citep{2006ApJ653593S}. Finally, we note that for nearby GF-candidate GRBs with unknown distance, the power of the scaling (Eq. \ref{eq:Ep_flux}) relation discussed in this letter is that one can use the relation between the peak energy and flux for distinguishing it from short GRBs. 

\section*{Data availability}
The data underlying this article will be shared on reasonable request to the corresponding author.

\section*{acknowledgments}
We thank A.R. Rao, K. Ioka, and XY Wang for discussions. BBZ acknowledges the supported by the Fundamental Research Funds for the Central Universities (14380035). This work is supported by National Key Research and Development Programs of China (2018YFA0404204), the National Natural Science Foundation of China (Grant Nos. 11833003, U1838105, U1831135) and the Program for Innovative Talents, Entrepreneur in Jiangsu, and the Strategic Priority Research Program on Space Science, the Chinese Academy of Sciences, Grant No. XDB23040400. RG and SBP acknowledge BRICS grant {DST/IMRCD/BRICS/PilotCall1/ProFCheap/2017(G)} for the financial support. 

Softwares used: {Astropy \citep{Astropy}, Emcee \citep{emcee}
}

\clearpage
\bibliographystyle{aasjournal}
\bibliography{msRAA-2020-0441}

\end{document}